# RATE DISTORTION STUDY FOR TIME-VARYING AUTOREGRESSIVE GAUSSIAN PROCESS

Jia-Chyi Wu

Department of Communications, Control and Navigation Engineering,
National Taiwan Ocean University, Keelung, Taiwan, ROC
jcwu@ntou.edu.tw

## ABSTRACT

*Formulation of the Rate-distortion association is an information-theoretic study in the field of signal encoding systems. Since a more general approach to model the nonstationarity exhibited by real-world signals is to use appropriately fitted time varying autoregressive (TVAR) models, we have investigated the rate-distortion function R(D) for the class of time varying nonstationary signals. In this study, we present formulations of the rate-distortion function for the Gaussian TVAR processes. The R(D) function can serve as an information-theoretic bound on the performance achievable by source encoding techniques when the processing signal is represented exclusively by a Gaussian TVAR model.*

## 1. INTRODUCTION

Rate-distortion analysis was first suggested in Shannon's original development of information theory [1] and it was exclusively developed for memoryless and Markov sources in conjunction with the squared-error distortion measure [2]. It was later generalized to stationary ergodic processes with discrete alphabets and to Gaussian processes by Gallager [3], as well as to stationary and abstract alphabets by Berger [4]. Many contributions have been made to rate-distortion theory for varied stationary sources [5]–[6], [17]–[23]. The theory is comprehensively discussed in books by Berger [4] and Gallager [3]. Extensive bibliographies for this subject appear in [7]–[9]. Recently, Jesús Gutiérrez-Gutiérrez, et. al. present an integral formula for the rate-distortion function of asymptotically wide sense stationary (AWSS) Gaussian vector process as well as the *R*(*D*) function of moving average (MA) Gaussian vector processes and of autoregressive MA (ARMA) AWSS Gaussian vector processes [24]–[25]. Due to the nonstationary characteristics of real-world signal sources, recent research interest has focused on extending the *R*(*D*) function for the nonstationary process. Berger [10] proved a source coding theorem with respect to a squared-error distortion measure for the Wiener process, a special case of the nonstationary Gaussian AR process. Gray [11] showed that the source coding theorem generally holds for the Gaussian AR process even if the process is neither stationary nor asymptotically stationary. Hashimoto and Arimoto [12] then developed a general form for the *R*(*D*) function for the class of a nonstationary Gaussian AR process wherein the variance of the process grows exponentially. However, the nonstationary Gaussian AR process indicated in these papers [10]–[12] becomes unstable asymptotically, since the variance of the AR process grows either exponentially or algebraically. Since the nonstationary characteristics exhibited by real-world signals can be modelled using appropriately fitted time varying autoregressive (TVAR) forms [13]–[14], we have investigated the *R*(*D*) formulation for the class of time varying nonstationary signals in this study.

In this paper we present the *R*(*D*) function study for Gaussian TVAR processes. First of all, we briefly review the *R*(*D*) function for time-discrete sources that are input to a general communication system. We then present the methodology to form the *R*(*D*) function for Gaussian TVAR processes. The *R*(*D*) function is an ultimate bound in assessing both the absolute and relative performance achievable by time varying signal source coding system.

## 2. PRELIMINARY REVIEW FOR *R*(*D*) FUNCTION

The rate-distortion function, *R*(*D*), is mathematically defined in terms of the average mutual information between the source outputs and the reproduced source outputs at the destination. Consider a time-discrete, continuous-amplitude, stationary source $\{x_t, t = \ldots, -1, 0, 1, \ldots\}$ having a joint probability density $P(\mathbf{x}) = P(x_1, x_2, \ldots, x_N)$ governing the generation of *N* successive source letters. Also, specify a single-letter fidelity criterion *F*, which is the distortion between a source word *x* and a reproducing word *y* present at the destination, to be defined as the arithmetic average of the distortions between the corresponding letters of **x** and **y** as assigned by a fixed single-letter nonnegative distortion measure, $\rho(x_t, y_t)$. Consider all possible conditional probability densities $Q(\mathbf{y}|\mathbf{x})$ associated with reproducing alphabet $\mathbf{y} = \{y_1, y_2, \ldots, y_N\}$ and the given source alphabet $\mathbf{x} = \{x_1, x_2, \ldots, x_N\}$.

The *R*(*D*) function for the particular source $\{x_t\}$ with respect to $F_\rho$ is defined as

$$R(D) = \lim_{N \to \infty} R_N(D), \tag{1}$$

$$R_N(D) = \frac{1}{N} \inf_{Q \in Q_D} I(Q), \tag{2}$$

where *I*(*Q*) is the average mutual information, defined as a function of the conditional probability density function. The limit in Eq. (1) always exists for a stationary source [4] so that *R*(*D*) is well-defined. In the particular case of a one-dimensional time-discrete stationary Gaussian source with power spectral density (PSD)

$$S(\omega) = \sum_{k=-\infty}^{\infty} c_k e^{-jk\omega}, \tag{3}$$

where $c_k$ is the autocorrelation function, and using the asymptotic form of *R*(*D*) in Eq. (1), it can be shown that the mean-squared error (MSE) rate-distortion function has the parametric representation [4],

$$D_\theta = \frac{1}{2\pi} \int_{-\pi}^{\pi} \min[\theta, S(\omega)] d\omega, \tag{4}$$

$$R(D_\theta) = \frac{1}{2\pi} \int_{-\pi}^{\pi} \max\left[0, \frac{1}{2} \log\left(\frac{S(\omega)}{\theta}\right)\right] d\omega, \tag{5}$$

with the non-zero portion of the *R*(*D*) curve being generated by the parameter $\theta$ in the interval $0 \le \theta \le \text{ess sup } S(\omega)$, where the arithmetic average squared error is defined as $(y_t - x_t)^2$. The *R*(*D*) function for a Gaussian source with memory is completely specified in terms of the source properties, namely the source power spectral density.

## 3. *R*(*D*) FUNCTION FOR GAUSSIAN TVAR PROCESS

The Gaussian time-varying AR process is of primary interest because of its wide applicability to the modeling of real-world sources, e.g., speech signals and rasterized image signals. In this section, we consider the one-dimensional Gaussian TVAR process. Let $\{x_t\}$ be a Gaussian TVAR process satisfying the difference equation

$$x_t = -\sum_{m=1}^{M} a_m\left(\tfrac{t}{N}\right) x_{t-m} + z_t, \text{ for } t = 1, 2, \ldots, N, \tag{6}$$

where $x_t = 0$ for $t = 0, -1, -2, \ldots$, and $z_t$ are independent and identically distributed (i.i.d.) zero-mean Gaussian random variables possessing variance $\sigma_z^2$. The terms $a_m\left(\frac{t}{N}\right)$ are the AR parameters varying with time $t = 1, 2, \ldots, N$. The time index $\left(\frac{t}{N}\right)$ of the time-varying AR parameters $a_m\left(\frac{t}{N}\right)$ is defined such that $0 < \left(\frac{t}{N}\right) \leq 1$.

Upon iterating Eq. (6) to obtain $\mathbf{x} = \{x_1, x_2, \ldots, x_N\}$, it is observed that $\mathbf{x}$ is specified uniquely by $\mathbf{z} = \{z_1, z_2, \ldots, z_N\}$ and the known initial state $\mathbf{x}_0 = \{x_0, x_{-1}, \ldots, x_{-m+1}\} = (0, 0, \ldots, 0)$. Therefore, $z_t$ can be specified uniquely by $\mathbf{x}$ and $\mathbf{x}_0$ via the relation

$$z_t = \sum_{m=0}^{M} a_m\left(\tfrac{t}{N}\right) x_{t-m}, \text{ for } t = 1, 2, \ldots, N, \qquad (7)$$

with $a_0(.) = 1$. Equation (7) can be represented in matrix form, $\mathbf{z} = \mathbf{A}\mathbf{x}$, for $t = 1, 2, \ldots, N$, where $\mathbf{A}$ is the $N \times N$ lower triangular matrix,

$$\mathbf{A} = \begin{bmatrix} a_0\left(\tfrac{1}{N}\right) & 0 & 0 & \cdots & \cdots & \cdots & 0 \\ a_1\left(\tfrac{2}{N}\right) & a_0\left(\tfrac{2}{N}\right) & 0 & \cdots & \cdots & \cdots & 0 \\ \vdots & \vdots & \ddots & \ddots & \vdots & \vdots & \vdots \\ a_M\left(\tfrac{M+1}{N}\right) & a_{M-1}\left(\tfrac{M+1}{N}\right) & \cdots & a_0\left(\tfrac{M+1}{N}\right) & 0 & \cdots & 0 \\ 0 & a_M\left(\tfrac{M+2}{N}\right) & \cdots & \cdots & a_0\left(\tfrac{M+2}{N}\right) & \cdots & 0 \\ \vdots & \vdots & \vdots & \vdots & \vdots & \vdots & \vdots \\ 0 & \vdots & 0 & a_M\left(\tfrac{N-1}{N}\right) & \cdots & a_0\left(\tfrac{N-1}{N}\right) & 0 \\ 0 & \cdots & \cdots & 0 & a_M(1) & \cdots & a_0(1) \end{bmatrix}. \qquad (8)$$

For the Gaussian AR process $\{x_t\}$, let $R_N(D)$ be the per letter rate-distortion function of the $n$-dimensional vector $(x_1, x_2, \ldots, x_N)$. The rate-distortion function of the process $\{x_t\}$ is defined as

$$R(D) = \lim_{N \to \infty} R_N(D), \qquad (9)$$

and from Berger [4] (6.3.34) and (6.3.35), $R_N(D)$ has the following parametric form when $x_t$ is an $N$-dimensional zero-mean Gaussian distributed random vector

$$\begin{aligned} D_\theta &= \frac{1}{N} \sum_{m=0}^{N} \min\left(\theta, \frac{1}{\alpha_m}\right), \\ R_N(D_\theta) &= \frac{1}{N} \sum_{m=0}^{N} \max\left(0, \frac{1}{2} \log \frac{1}{\theta \alpha_m}\right), \end{aligned} \qquad (10)$$

where, $\alpha_m$ are the eigenvalues of the inverse autocorrelation matrix $\boldsymbol{\Phi}_N^{-1}$ of the random variables $x_1, x_2, \ldots, x_N$, and $\theta$ is a parameter taking on values in the interval $0 \leq \theta < \alpha_{max}$. From the relationship $\mathbf{z} = \mathbf{A}\mathbf{x}$, we then have

$$\boldsymbol{\Phi}_N = E[\mathbf{X}\mathbf{X}^T] = \mathbf{A}^{-1} E[\mathbf{z}\mathbf{z}^T] (\mathbf{A}^T)^{-1} = \sigma_z^2 (\mathbf{A}^T \mathbf{A})^{-1}, \qquad (11)$$

and $\mathbf{A}^{-1}$ exists because

$$\det \mathbf{A} = \prod_{t=1}^{N} a_0\left(\tfrac{t}{N}\right) = 1 \neq 0. \qquad (12)$$

Hence, the inverse autocorrelation matrix is found to be

$$\mathbf{\Phi}_N^{-1} = \left(\frac{1}{\sigma_z^2}\right)\mathbf{A}^T\mathbf{A}. \tag{13}$$

The entries in the inverse matrix $\mathbf{\Phi}_N^{-1}$ are given as $\phi_N^{-1}(\mu,\nu)$, and

$$\phi_N^{-1}(\mu,\nu) = \frac{1}{\sigma_z^2}\sum_{m=0}^{N-\max(\mu,\nu)} a_m\left(\frac{m+\max(\mu,\nu)}{N}\right) a_{m+|\mu-\nu|}\left(\frac{m+\max(\mu,\nu)}{N}\right).$$

Since $a_m(.) = 0$ for $m > M$, it follows that all entries more than $M$ diagonals away from the main diagonal of $\mathbf{\Phi}_N^{-1}$ are zero. Also, if either $\mu \leq N{-}M$ or $\nu \leq N{-}M$, then $\phi_N^{-1}(\mu,\nu) = \phi_N^{-1}(|\mu-\nu|)$, and

$$\phi_N^{-1}(|\mu-\nu|) = \frac{1}{\sigma_z^2}\sum_{m=0}^{M} a_m\left(\frac{m+\max(\mu,\nu)}{N}\right) a_{m+|\mu-\nu|}\left(\frac{m+\max(\mu,\nu)}{N}\right). \tag{14}$$

Hence $\mathbf{\Phi}_N^{-1}$ is a Hermitian matrix. Due to the Hermitian structure of $\mathbf{\Phi}_N^{-1}$, it is possible to consider the limiting case of infinite $N$ in Eq. (10). By invoking a theorem defined by Grenander and Szegö [15] regarding the asymptotic distribution of the eigenvalues of certain Hermitian structures, we first consider the class of real-valued functions $f(r, \omega)$, $0 \leq r \leq 1$, periodic in $\omega$ with period $2\pi$, satisfying the following condition,

*Condition:*

The coefficients $\psi_n(r)$ of the Fourier series

$$f(r,\omega) = \sum_{n=-\infty}^{\infty}\psi_n(r)e^{-jn\omega} \tag{15}$$

are continuous and there exists a constant $K$ such that

$$\sum_{n=-\infty}^{\infty}\max|\psi_n(r)| \leq K, \tag{16}$$

where the maximum values are taken in the interval $0 \leq r \leq 1$.

The Fourier series of $f(r, \omega)$ is absolutely convergent and the function $f(r, \omega)$ of the two variables $r$ and $\omega$ is continuous. We then have the following theorem,

**Theorem I. Asymptotic distribution theorem of the eigenvalues of the specified Hermitian form** (ref. Grenander and Szegö [15], Theorem 6.5)

Let the function $f(r, \omega)$, where $r = \frac{\max(\mu,\nu)}{N}$, satisfy the above condition. We denote the eigenvalues of the Hermitian form

$$\sum_{\mu,\nu}\psi_{\nu-\mu}\left(\frac{\max(\mu,\nu)}{N}\right)u_\mu u_\nu, \quad \mu, \nu = 0, 1, 2, \ldots, N. \tag{17}$$

by $\lambda_\nu^{(N)}$, we have then $|\lambda_\nu^{(N)}| \leq K$. Moreover, if $F(\lambda)$ is any continuous function defined for $-K \leq \lambda \leq K$, we have

$$\lim_{N\to\infty}\frac{1}{N}\sum_{\nu=1}^{N}F(\lambda_\nu^{(N)})=\frac{1}{2\pi}\int_{-\pi}^{\pi}\int_0^1 F[f(r,\omega)]drd\omega. \tag{18}$$

In the special case when $f(r, \omega)$ is independent of $r$, this is an assertion on Toeplitz forms. The limit relation Eq. (17) is equivalent to the following special case [15],

$$\lim_{N\to\infty}\frac{1}{N}\sum_{\nu=1}^{N}(\lambda_\nu^{(N)})^k=\frac{1}{2\pi}\int_{-\pi}^{\pi}\int_0^1 [f(r,\omega)]^k drd\omega, \quad k=0,1,2,... \tag{19}$$

Theorem I can be attested directly from the procedures described both in Grenander and Szegö [15] Theorem 1.18(b) and Theorem 6.5.

Since we have defined $r=\frac{\max(\mu,\nu)}{N}$, $\mu, \nu = 1, 2, ..., N$, $\phi_N^{-1}(\mu,\nu)$ is rewritten as

$$\phi_N^{-1}(|\mu-\nu|)=\phi_{|\mu-\nu|}^{-1}(r)=\frac{1}{\sigma_z^2}\sum_{m=0}^{M}a_m\left(\frac{m}{N}+r\right)a_{m+|\mu-\nu|}\left(\frac{m}{N}+r\right). \tag{20}$$

Now, let $\mathbf{G}_N = \mathbf{\Phi}_N^{-1}$ and $\{\mathbf{G}_N\}$ be sequences of Hermitian matrices with eigenvalues $\{\alpha_m^{(N)}\}$. The entries of the symmetric Hermitian matrix $\mathbf{G}_N$ are

$$g_{|\mu-\nu|}(r)=g_k(r)=\frac{1}{\sigma_z^2}\sum_{m=0}^{M}a_m\left(\frac{m}{N}+r\right)a_{m+|\mu-\nu|}\left(\frac{m}{N}+r\right) \tag{21}$$

on the $k^{th}$ diagonal. In most real-world signal processing cases, the autoregressive order $M$ is much less than $N$, therefore, we have that $m/N \to 0$ when $N \to \infty$. Equation (21) is simplified as

$$g_k(r)=\frac{1}{\sigma_z^2}\sum_{m=0}^{M}a_m(r)a_{m+k}(r). \tag{22}$$

Since the $\mathbf{G}_N$ possess the Hermitian form specified by Eq. (17), Theorem I implies that their eigenvalues $\{\alpha_m^{(N)}\}$ are distributed asymptotically according to

$$g(r,\omega)=\sum_{k=-\infty}^{\infty}g_k(r)e^{-jk\omega}=\frac{1}{\sigma_z^2}\left|\sum_{m=0}^{M}a_m(r)e^{-jm\omega}\right|^2 \tag{23}$$

with $\omega$ uniform on $[-\pi, \pi]$, and that

$$\lim_{N\to\infty}|\mathbf{G}_N|=\sqrt{\frac{1}{2\pi}\int_{-\pi}^{\pi}\int_0^1 g^2(r,\omega)drd\omega}<\infty, \tag{24}$$

where $|\mathbf{G}_N|$ represents the weak norm of $\mathbf{G}_N$,

$$|\mathbf{G}_N|=\sqrt{\frac{1}{N}\sum_{\mu,\nu=1}^{N}|G_N(\mu,\nu)|^2}=\sqrt{\frac{1}{N}\sum_{m=1}^{N}(\alpha_m^{(N)})^2}. \tag{25}$$

The integral in Eq. (24) is finite due to the fact that

$$0\leq g(r,\omega)\leq\frac{1}{\sigma_z^2}\left(\sum_{m=0}^{M}|a_m(r)|\right)^2<\infty. \tag{26}$$

Now, by applying Theorem I, we have that

$$\lim_{N\to\infty}\frac{1}{N}\sum_{m=1}^{N}F(\alpha_m^{(N)})=\frac{1}{2\pi}\int_{-\pi}^{\pi}\int_0^1 F[g(r,\omega)]drd\omega \tag{27}$$

for any continuous function $F$ on $[\delta, \mu]$, where $\delta$ and $\mu$ are the essential infimum and supremum of $g(r, \omega)$, respectively. Applying Eq. (27) so as to pass to the limit in Eq. (10), we obtain the following theorem as a consequence.

**Theorem II. Rate-distortion function for time-varying autoregressive Gaussian process**

Let $x_t$ be an $M^{th}$-order Gaussian TVAR source generated by an i.i.d. $N(0, \sigma_z^2)$ sequence $z_t$ and the TVAR coefficients $a_m(r)$, $m = 1, 2, ..., M$, where $r = \frac{t}{N}$. Then the mean-squared error (MSE) rate-distortion function of $x_t$ is given parametrically by

$$D_\theta = \frac{1}{2\pi} \int_{-\pi}^{\pi} \int_0^1 \min\left[\theta, \frac{1}{g(r,\omega)}\right] dr d\omega,$$

$$R(D_\theta) = \frac{1}{2\pi} \int_{-\pi}^{\pi} \int_0^1 \max\left[0, \frac{1}{2}\log\frac{1}{\theta g(r,\omega)}\right] dr d\omega,$$
(28)

where

$$g(r,\omega) = \frac{1}{\sigma_z^2}\left|1 + \sum_{m=1}^{M} a_m(r) e^{-jm\omega}\right|^2.$$
(29)

## 4. SUMMARY AND CONCLUSIONS

The rate-distortion function $R(D)$ for time-varying autoregressive (TVAR) nonstationary signals, based upon the theorem from Grenander and Szegö [15] regarding the asymptotic distribution of the eigenvalues of certain Hermitian forms, is investigated and formulated in this study. The $R(D)$ function is served as an information-theoretic bound on the performance achievable by source encoding techniques when the processing signal is represented exclusively by a Gaussian TVAR model. The rate-distortion function can be used as an ultimate performance bound in assessing both the absolute and relative performance achievable by any specific time-varying signal processing system.

## ACKNOWLEDGEMENTS

This work has been supported partially by the Ministry of Science and Technology in Taiwan, the Republic of China under grant number MOST 107-2119-M-004.

## REFERENCES

[1] C. E. Shannon, (1948) "A mathematical theory of communication," Bell Syst. Tech. Journal, vol. 27, pp. 379–423 and pp. 623–656.

[2] C. E. Shannon, (1959) "Coding theorems for a discrete source with a fidelity criterion," IRE Natl. Cov. Rec., Part 4, pp. 142–163.

[3] R. G. Gallagher, (1968) *Information Theory and Reliable Communications*. New York: Wiley & Sons, ch. 9.

[4] T. Berger, (1971) *Rate Distortion Theory: A Mathematical Basis for Data Compression*. Englewood Cliffs, NJ: Prentice Hall.

[5] R. E. Blahut, (1972) "Computation of channel capacity and rate-distortion functions," IEEE Trans. on Inform. Theory, vol. IT-18, pp. 460–473.

[6] L. D. Davisson, (1972) "Rate-distortion theory and application," Proceedings of IEEE, vol. 60, pp. 800–808.

[7] H. C. Andrews, (1971) "Bibliography on rate distortion theory," IEEE Trans. on Inform. Theory, vol. IT-17, pp. 198–199.

[8] L. C. Wilkins and P. A. Wintz, (1971) "Bibliography on data compression, picture properties, and picture coding," IEEE Trans. on Inform. Theory, vol. IT-17, pp. 180–197.

[9] J. C. Kieffer, (1993) "A survey of the theory of source coding with a fildelity criterion," IEEE Trans. on Inform. Theory, vol. IT-39, pp. 1473–1490.

[10] T. Berger, (1970) "Information rates of Wiener processes," IEEE Trans. on Inform. Theory, vol. IT-16, pp. 134–139.

[11] R. M. Gray, (1970) "Information rates of autoregressive processes," IEEE Trans. on Inform. Theory, vol. IT-16, pp. 412–421.

[12] T. Hashimoto and S. Arimoto, (1980) "On the rate-distortion function for the nonstationary Gaussian autoregressive process," IEEE Trans. on Inform. Theory, vol. IT-26, pp. 478–480.

[13] G. Alengrin, M. Barlaud, and J. Menez, (1986) "Unbiased parameter estimation of nonstationary signals in noise," IEEE Trans. on Acoust., Speech, Signal Processing, vol. 34, pp. 319–1322.

[14] Y. Grenier, (1983) "Time-dependent ARMA modeling of nonstationary signals," IEEE Trans. on Acoust., Speech, Signal Processing, vol. ASSP-31, pp. 899–911.

[15] U. Grenander and G. Szegö, (1958) *Toeplitz Forms and Their Applications*, CA: Univ. of Cal. Press.

[16] Y. Steinberg and S. Verdú, (1996) "Simulation of random processes and rate-distortion theory," IEEE Trans. on Inform. Theory, vol. IT-42, pp. 63–86.

[17] J. Muramatsu and F. Kanaya, (1994) "Distortion-complexity and rate-distortion function," IEICE Trans. Fundamentals, Vol.E77-A, No.8, pp.1224–1229.

[18] F. Kanaya and J. Muramatsu, (1997) "An almost sure recurrence theorem with distortion for stationary ergodic sources," IEICE Trans. Fundamentals, E80-A, No.11, pp.2264–2267.

[19] K. Iwata and J. Muramatsu, (2002) "An information-spectrum approach to rate-distortion function with side information," IEICE Trans. Fundamentals, Vol.E85-A, No.6, pp.1387–1395.

[20] T. M. Cover and M. Chiang, (2002) "Duality between channel capacity and rate distortion with two-sided state information," IEEE Trans. on Inform. Theory, vol. IT-48, pp. 1629–1638.

[21] S. Cheng, V. Stankovic, Z. Xiong, (2005) "Computing the channel capacity and rate-distortion function with two-sided state information," IEEE Trans. on Info. Theory, vol. 51, pp. 4418–4425.

[22] M. Fleming and M. Effros, (2006) "On the rate-distortion with mixed types of side information", IEEE Trans. on Inform. Theory, vol. IT-52, pp. 1698–1705.

[23] J. Binia and H. Israel, (2008) "Rate distortion function with a proportional mean-square error distortion measure," in Proceedings of the International Symposium on Information Theory (ISIT), pp. 862-866, Toronto, Canada.

[24] Gutiérrez-Gutiérrez, J.; Zárraga-Rodríguez, M.; Villar-Rosety, F.M.; Insausti, X., (2018) "Rate-Distortion Function Upper Bounds for Gaussian Vectors and Their Applications in Coding AR Sources," Entropy, Vol. 20, 399.

[25] Gibson, J., (2017) "Rate Distortion Functions and Rate Distortion Function Lower Bounds for Real-World Sources" Entropy, Vol. 19, 604.